\documentclass[pre,superscriptaddress]{revtex4}

\usepackage{graphicx}
\usepackage{subfigure}

\newcommand{\figsubwidth}{5.8cm}

\begin{document}

\markboth{\"Ozg\"ur, Cetin, Bingol}
{Co-occurrence Network of Reuters News}


\title{Co-occurrence Network of Reuters News}

\author{Arzucan \"Ozg\"ur}
\address{Department of Electrical Engineering and Computer Science, University of Michigan, 
Ann Arbor, MI, USA }

\author{Burak Cetin}
\address{Complex Systems Research Laboratory, 
Department of Computer Engineering, 
Bogazici University 
Istanbul, Turkey }

\author{Haluk Bingol}
\address{Complex Systems Research Laboratory, 
Department of Computer Engineering, 
Bogazici University 
Istanbul, Turkey }


\begin{abstract}
Networks describe various complex natural systems including social systems. 
We investigate the social network of co-occurrence in Reuters-21578 corpus, 
which consists of news articles that appeared in the Reuters newswire in 1987. 
People are represented as vertices and two persons are connected if they co-occur in the same article.
The network has small-world features with power-law degree distribution. 
The network is disconnected and the component size distribution has power law characteristics. 
Community detection on a degree-reduced network provides meaningful communities.
An edge-reduced network, which contains only the strong ties has a star topology.

``Importance'' of persons are investigated.
The network is the situation in 1987.
After 20 years, a better judgment on the importance of the people can be done.
A number of ranking algorithms, including Citation count, PageRank, are used to assign ranks to vertices.
The ranks given by the algorithms are compared against how well a person is represented in Wikipedia. 
We find up to medium level Spearman's rank correlations. 
A noteworthy finding is that PageRank consistently performed worse than the other algorithms.
We analyze this further and find reasons.

\end{abstract}

\maketitle

\section{Introduction}
Networks describe various complex real world systems such as neural network of a worm (Caenorhabditis elegans), power grid of the Western United States, phone call networks, networks of linguistics, protein-folding, the World Wide Web and social systems~\cite{Newman2003}. 
Social networks describe human societies whose nodes are individual people and links represent a social interaction among these people. 
Although obtaining data about social networks is difficult, many interesting social networks such as 
social network of scientific collaboration~\cite{Barabasi02}, 
movie actors' collaboration~\cite{Watts98}, 
sexual contacts~\cite{Liljeros01}, 
email lists~\cite{Kirlidog03}, 
and terrorist network of September 11, 2001~\cite{Krebs02} 
have been studied in the literature. 
Recent studies have shown that these networks possess some common properties such as being small-world and scale-free. 

Reuters-21578 corpus is a standard data set used extensively in research in automatic document categorization, information retrieval, machine learning and other corpus-based research~\cite{ReutersCorpus}. 
It consists of $21,578$ news articles, mostly about economics, that appeared in the Reuters newswire in 1987. 
In a previous study, 
$3,000$ of these news articles are randomly selected and person names are manually identified by reading~\cite{Ozgur2004}. 
An undirected, un-weighted network where nodes are people and there is a link if two people have appeared in the same news article is constructed. 
Then, the scale-free and small world properties of the network is studied. 

In this study we extend the network to cover all the $21,578$ news articles.
We investigate component structure, community structure, co-occurence in time, and the ``importance'' of people.

\section{Construction of The Social Network}

The most challenging part of the study was constructing the network~\cite{ReutersNetwork}, since it required a lot of manual effort. 
An undirected and weighted graph of social network of co-occurrence in news articles is constructed as follows:
\begin{enumerate}
\item All the news articles in the Reuters-21578 corpus  are processed and person names are identified. 
Person names are identified manually in $10,000$ of the articles. 
To identify the names in the remaining articles, we used the name finder tool of Bikel et al.~\cite{Bikel97}. 
The output of the tool is post-processed manually to eliminate the errors. 
\item Nodes of the network are defined as distinct people.
\item A link is constructed between two people if their names appear in the same news article. Each edge is associated with a weight corresponding to the number of times the two people appeared in the news stories together.
\end{enumerate}

Note that in this representation there is an edge in between two persons if they co-occur in one article independent of the number of co-occurrence. 
Therefore, the graph represents connectivity but it could be misleading in terms of ``closeness'' of the people. Another observation about the representation is that the degree of a vertex could be misleading. 
Suppose there are 5 people in an article which contribute 4 degrees to each corresponding vertex in the graph. Probably this situation is quite different than 5 persons co-occurring pairwise in different articles although the graph representation is the same for both cases. The vertex is not weighted. Either a person is in one single article or she is in many articles, the representation is the same. The social network constructed consists of $5,249$ nodes and $7,528$ edges. 

\section{Connected Components}
The social network of co-occurrence in news articles turns out to be an unconnected graph. 
The graph consists of $2,105$ components. 
There is a giant component of $1,682$ vertices and $5,008$ edges. 
The rest of the components are small with sizes in the range of [1, 26]. 
Table \ref{comp} displays component size, 
number of occurrences of that component size in the network, number and percentage of people who appear in a component of the specified size.

\begin{table}[htp]
\caption{Component size frequency table.}
\begin{tabular}{ccccc} 
\toprule
Size $s$ & Frequency $p(s)$ & \% Frequency & \# of Persons & \% of Persons \\
\colrule
\hphantom{000}1		& 1363				& 64.75 			& 1363 				& 25.97\\
\hphantom{000}2		& \hphantom{0}442	& 20.42 			& \hphantom{0}884	& 16.84 \\
\hphantom{000}3		& \hphantom{0}138	& \hphantom{0}6.56	& \hphantom{0}414	& \hphantom{0}7.89 \\
\hphantom{000}4 	& \hphantom{00}71  	& \hphantom{0}3.37 	& \hphantom{0}284 	& \hphantom{0}5.41 \\
\hphantom{000}5 	& \hphantom{00}35  	& \hphantom{0}1.66 	& \hphantom{0}175 	& \hphantom{0}3.33 \\
\hphantom{000}6 	& \hphantom{00}19  	& \hphantom{0}0.90 	& \hphantom{0}114 	& \hphantom{0}2.17 \\
\hphantom{000}7 	& \hphantom{00}11  	& \hphantom{0}0.52 	& \hphantom{00}77 	& \hphantom{0}1.47 \\
\hphantom{000}8 	& \hphantom{000}8  	& \hphantom{0}0.38 	& \hphantom{00}64 	& \hphantom{0}1.22 \\
\hphantom{000}9 	& \hphantom{000}5	& \hphantom{0}0.24 	& \hphantom{00}45 	& \hphantom{0}0.86 \\
\hphantom{00}10 	& \hphantom{000}5 	& \hphantom{0}0.24 	& \hphantom{00}50 	& \hphantom{0}0.95 \\
\hphantom{00}11 	& \hphantom{000}3  	& \hphantom{0}0.14 	& \hphantom{00}33 	& \hphantom{0}0.63 \\
\hphantom{00}12  	& \hphantom{000}2  	& \hphantom{0}0.10 	& \hphantom{00}24 	& \hphantom{0}0.46 \\
\hphantom{00}14  	& \hphantom{000}1  	& \hphantom{0}0.05 	& \hphantom{00}14 	& \hphantom{0}0.27 \\
\hphantom{00}26 	& \hphantom{000}1  	& \hphantom{0}0.05 	& \hphantom{00}26 	& \hphantom{0}0.50 \\
1682 				& \hphantom{000}1	& \hphantom{0}0.05	& 1682				& 32.04 \\
\botrule
\end{tabular}
\label{comp}

\end{table}

$64.75\%$ of the components are of size $1$. 
This means that $1,363$ people in the network are not connected to anyone, 
i.e., they appear just by themselves in the news. 
These people are generally US senators, and managers or chairs of certain corporations. 
In the news articles they appeared by expressing their ideas about a phenomenon or by being elected for a certain position. 
Examples of such patterns are ``person A told'' and ``person A was elected to''. 
There are $442$ components of size $2$. In other words, there are $442$ pairs of people who are connected only to each other. Patterns of news that form components of size 2 are as ``person A said and person B said'' and ``person A was chosen to replace person B''. All of the components of size less than and equal to $26$ possess similar characteristics.
  
We investigate the frequency $p(s)$ of occurrence of components of size $s$. 
We observed that $p(s)$ is related to $s$ by a power law given by 
$p(s) \propto s^{- \gamma}$, where $\gamma = 2.8$.

In the rest of the study, the results are computed based on the giant component.

\section{Complex Network Analysis}
The diameter, that is, the largest distance between two nodes, of the network is calculated to be $13$. 
This is between \textit{Jack Sandner} (Chairman of Chicago Mercantile Exchange) 
and \textit{Richard Simon} (analyst at Goldman Sachs and Co). 
Interestingly, these people are both from USA. 
So, although they are geographically close to each other, they are the most distant people in the network of news. The diameter of a random graph with the same number of nodes and edges is $15$. 

The average path length and clustering coefficient are given in Table~\ref{swmetrics}. 
They are calculated by Pajek~\cite{Pajek}.
The diameter and average path length of the network are relatively small compared to the size of the network. 
The clustering coefficient is $\sim 474$ times greater than the clustering coefficient of a random network with the same number of nodes and edges. 
Thus, we can conclude that this is very different from a random graph.

\begin{table}[ht]
\caption{Comparison of co-occurrence network with random network with the same number of vertices and edges.}
{\begin{tabular}{ccc} 
\toprule
Metric		& Co-occurrence Network & Random Network \\
\colrule
Diameter	&	13	&	15 \\
Average path length	& 4.39	& 6.73 \\
Clustering coefficient	& 0.6906	& 0.0014 \\
\botrule
\end{tabular} \label{swmetrics}}
\end{table}

The degree distribution of the network is power-law with exponent $\gamma = 2.4$. 
90.66\% of the nodes have degrees less that 10. 
The node with the maximum degree is the node representing Ronald Reagan 
which acts as the hub of the network since 23.50\% of the nodes are connected directly to it. 
Other highly connected nodes in decreasing order are given in Table~\ref{degreeTop}.

\begin{table}[ht]
\caption{Persons with the highest degree.}
{\begin{tabular}{ccc} 
\toprule
Person	 & Degree	& Note \\
\colrule
Ronald Reagan 		& 395 & President, USA \\
James Baker 		& 158 & Treasury Secretary, USA \\
Yasuhiro Nakasone 	& \hphantom{0}99 & Prime Minister, Japan \\
Margaret Thatcher 	& \hphantom{0}52 & Prime Minister, UK \\
George Bush 		& \hphantom{0}52 & Vice President, USA \\
Gerhard Stoltenberg & \hphantom{0}50 & Finance Minister, West Germany \\
William Casey 		& \hphantom{0}32 & Head of CIA, USA \\
Mikhail Gorbachev 	& \hphantom{0}28 & President, USSR \\
\botrule
\end{tabular} \label{degreeTop}}
\end{table}

\section{Communities}
Nodes in many real networks appear to group in sub-graphs where the density of connections among the nodes in the sub-graph is larger than that of the connections with the rest of the nodes in the network. 
These sub-graphs are called \textit{communities}. 
Since the community structure of the network is unknown, applying a community detection to this network would not be very useful since we cannot process the communities manually.
We decided to reduce the graph size by removing the vertices with degree less than 10 and obtained manageable graph of 49 people.
We use the Girvan-Newman betweenness clustering algorithm on the reduced network to detect the communities~\cite{Girvan02}. 
We identified four communities.
When we manually investigate the properties of the people in the communities (based on their job positions in 1987), we observe meaningful communities.

In the first community of 21 persons, we see US President Reagan, the prime minister of Japan, US senators, White House spokesmen, and news reporters. 
We also see that people involved in US national security are in this community. 
Current and former heads of CIA, Senate Armed Services Committee Chairman, current and former national security advisors, defense secretary of USA and two news reporters are clustered in this community. 
Interestingly, Marine Lt. Col. Oliver North, one of the main actors of ``Iran-Contra Affair''~\cite{irangate} political scandal, is also in this community.

There are 8 persons in the second community where we see US democratic senators, former president of USA Franklin Roosevelt, Secretary of the Commerce Department, and American Enterprise Institute analyst. 
It is interesting that we have a baseball manager in this community. 

In the third community, there are 9 persons many of which are related to Federal Reserve such as the former and current Federal Reserve board chairmen and vice chairman, staff director of monetary fund, New York Federal Reserve Bank President, and Federal Deposit Insurance Corp. Chairman. 
Former president of USA Jimmy Carter is also in this community. 

In the fourth community we see the Vice President of USA (later becomes president) George Bush, US democratic senators, commerce secretary, and secretary of state. The Canadian President and foreign secretary are also in this community.

\section{Strong Ties}
\begin{figure}
\includegraphics[width=\columnwidth,height=\columnwidth,keepaspectratio=true]{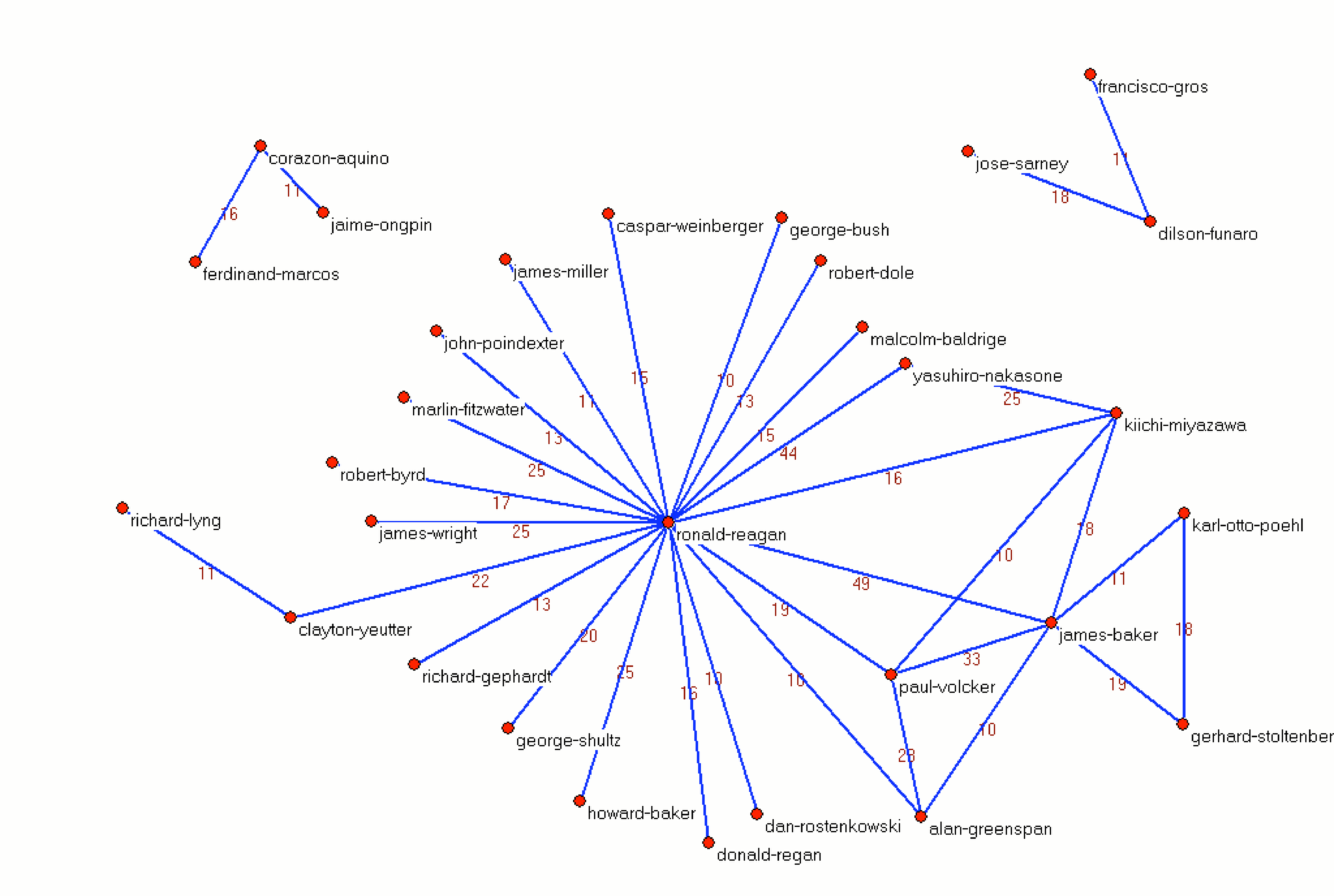}
\caption{Strong ties}
\label{fig:ties}
\end{figure}

We want to investigate the people connected strongly.
In order to manually check, we need to reduce the network size, again.
This time the edges with weight less than 10 are removed. 
The rationale is that if two people appear in only a few articles together they have a weaker relationship compared to people who often appear together therefore have high edge weight. 
So, we are left with edges that have high weight and associate these edges with representing strong relationships. 
Fig.~\ref{fig:ties} shows the cluster of the network. 
It is interesting to note that the topology is similar to a star connection, where Ronald Reagan is in the center. The highly weighted connections are between Ronald Reagan and other people in the White House such as Paul Volcker and James Baker; or presidents and politicians from other countries such as Yasuhiro Nakasone (President of Japan). 

There are two 4-cliques: 
\{Ronald Reagan, Paul Volcker, James Baker, Alan Greenspan\} and 
\{Ronald Reagan, Paul Volcker, James Baker, Kiichi Miyazawa\}. 
There are two triangles: 
\{Ronald Regan, Kiichi Miyazawa, Yasuhiro Nakasone\} and 
\{James Baker, Karl Otto Poehl, Gerhard Stoltenber\}.

We also note two triads that are unconnected from the star sub-network. 
The triad on the top left are people from Philippines 
\{President, former president, Finance Minister\} 
and the triad on the top right are people from Brazil 
\{President, Finance Minister, Chairman of a bank\}. 

\section{Time Series of Co-occurrence}
We used the chronological order property of the news articles to compute and visualize the patterns in the time-series of occurrences of people in the news. 
21,578 articles are chronologically sorted and binned into 43 bins of $500$ articles each based on their date. 

\begin{figure}[htbp]
\centering
\subfigure[] 
{
	\includegraphics[width=\figsubwidth,height=\figsubwidth,keepaspectratio=true]{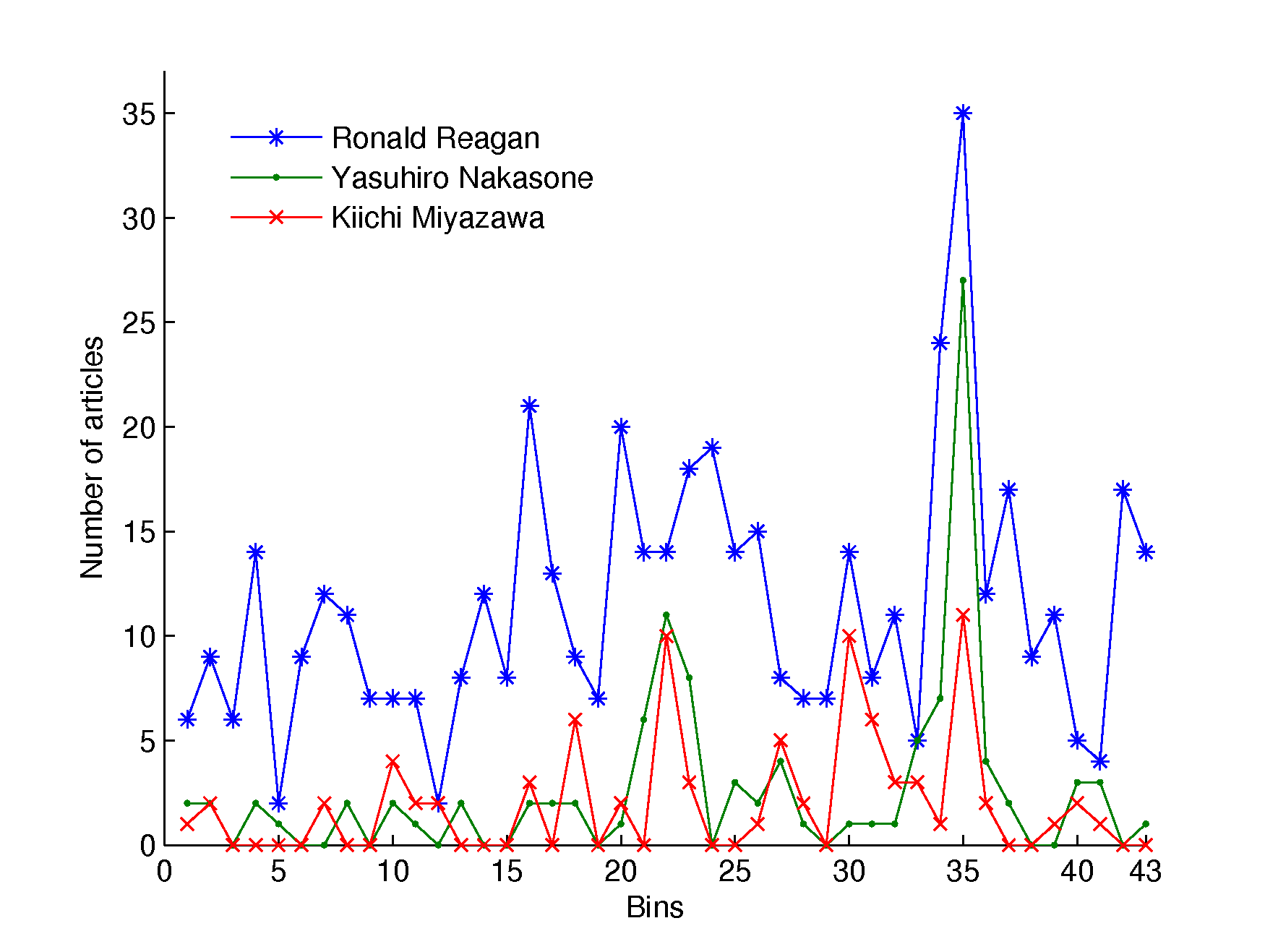}
	\label{fig:pattern1}
}
\hspace{.1in}
\subfigure[] 
{
	\includegraphics[width=\figsubwidth,height=\figsubwidth,keepaspectratio=true]{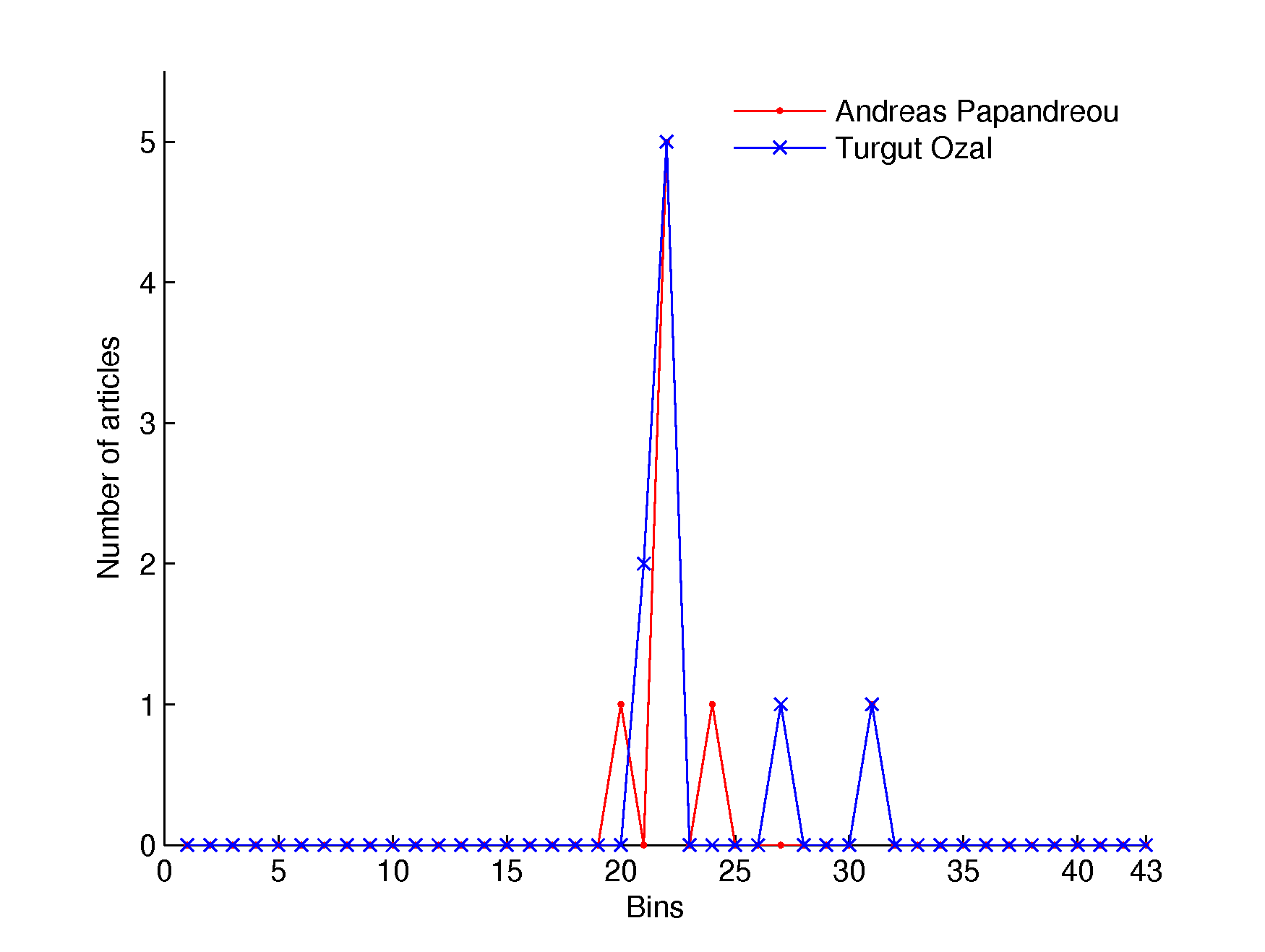}
	\label{fig:pattern2}
}
\caption{
The frequency of occurrence changes in time.
(a) The frequency of occurrence graphs of Ronald Reagan, Yasuhiro Nakasone and Kiichi Miyazawa present persistent continuous activation.
(b) The frequency of occurrence graphs of Turgut Ozal and Andreas Papandreou present partial activation based on some events.
}
\label{fig:frequencyOccurrence} 
\end{figure}

We graph the results. 
The x-axis, representing the time, is the bin number and the y-axis is the frequency of occurrence of the person in the news at the corresponding bin.  
The time series in 
Fig.~\ref{fig:frequencyOccurrence}~(a), 
shows the occurrence of Ronald Reagan (President of USA), Yasuhiro Nakasone (President of Japan), and Kiichi Miyazawa (Finance Minister of Japan). 
We note the frequency of occurrence of Ronald Reagan never drops to $0$. 
In general, what we observe is that, 
for popular people (presidents, politicians, and etc.) there is a continuous occurrence throughout the period. 
Some other people (maybe less popular in this sense) make a peak at some point, but later disappear and 
never re-occur again. 
There is an interesting peak at bin $35$. 
Due to space limitation the figure is not provided but 
when we look at some other people, James Baker and Paul Volcker also make a peak at that time point. 
On the other hand, people from other countries do not have peaks at that point. 
These patterns might show us that there is an important event going on between Japan and USA at that time point.

Another pattern that reveals an event between two countries is shown in Fig.~\ref{fig:frequencyOccurrence}~(b). 
Turgut Ozal is the president of Turkey and Andreas Papandreou is the president of Greece. At the time period that both presidents make peaks there was a crisis between the two countries.

\section{Finding Important Persons in the Network}
\label{sec:ImportantPersons}
The co-occurrence network contains information about the importance of people.
We investigate this by applying various vertex ranking measures on our network for a comparative study.
We expect to obtain the importance of persons as viewed from a Reuters perspective. 

%

\subsection{Analysis Setup}
In our study we will link the importance of a person in 1987 to importance today. 
We will see how well a person in the collection is represented in today's English Wikipedia \cite{Wikipedia} and compare that with rankings produced by various well-known centrality measures.
These are the algorithms:

\begin{enumerate}
	\item \textit{Article count}, denoted by \textit{acount}, is the number of articles a person appears in.
	\item \textit{Degree} is the number of people a person got associated with in the collection, i.e. the link count on the node.
	\item \textit{PageRank}, denoted by \textit{pr}, is the PageRank of a node as described in Page et. al.~\cite{Page1998}, using $d = 0.5$. For application we have converted the undirected network to directed by replacing each edge with arcs in both directions.
	\item \textit{Closeness}, denoted by \textit{cls}, calculated using the undirected unweighted network.
\textit{Closeness} calculates the shortest distance from each person in the network to each other person based on the links between all members of the network~\cite{FRE79}. 
Here central nodes are the ones which are closest to all other nodes.
	\item \textit{Betweennes}, denoted by \textit{btw}, calculated using the undirected unweighted network.
\textit{Betweenness} examines the extent to which a node is situated between others in a network~\cite{FRE79}.
It is a measure of how much damage to connectivity there would be if a given node is removed from the network.

\end{enumerate}

For assessing the validity of our results we have used a crawler to look up if a given person has an English Wikipedia page. 
We have interpreted this as an indication that a given person is important today in a general global sense. 
This would have an English speaking world bias and may not necessarily be a truly objective measure. 
However Reuters also being an English source and English being the closest there is to a truly global language, this measure should function at least to a reasonable extent. 
Our basic assertion here is that if a person was important back in 1987 when the Reuters articles were being published, then s/he would still be important today. 
The 20 years passed since then can make a ``time's judgment'' on who were truly important at the time. It is possible however other people in those articles unimportant or unforeseeable at the time will have gained importance. Similarly some who were not very important from a Reuters reporting perspective can actually be important individuals for different reasons.  
Combined, these would mean that the assessment power of the algorithms would be limited in discovering all those who are important, however this analysis should be reasonably good enough to penalize ``false positives'' which the algorithms would mark as important but were really not as such. 

Using the crawler results we have constructed two functions: 
\begin{enumerate}
	\item \emph{``has a page'':} $H(i) = 1$ if there is any Wikipedia page for a given person $i$. 
$H(i) = 0$ otherwise.
	\item \emph{``is a wikified page'':} $W(i) = 1$ if there is a Wikipedia page with the ``Wikified'' format for person $i$ which means the page is updated according to Wikipedia biography standards e.g. has a v-card biography box giving basic details. 
$W(i) = 0$, otherwise.
\end{enumerate}

Of the 5,249 persons in the network we find that 1,440 have a Wikipedia page and 383 have a ``Wikified'' page. In the rest of this section we will use these functions as apriori information on the importance of nodes and perform a comparative study of the algorithms. 
Table~\ref{tab:top20acount} shows the top 20 people when ranked according to $acount$ values. 
Having a glance at this table can serve as a basic reality check for the utility of our defined functions. 
For example we see that many people we could expect to have high importance have $W(i)=1$; President of USA, Prime Minister of Japan, Secretary of State of USA. 
Table~\ref{tab:top20closeness} shows a similar table for $cls$.

\begin{table}[hbt]
\caption{Top-20 persons in article count.}
{
	\begin{tabular}{@{}crccc@{}} 
	\toprule
	person & $acount$ & $H(i)$ & $W(i)$ & notes  \\
	\colrule
	ronald-reagan & 493 & 1 & 1 & President, USA \\
	james-baker & 212 & 1 & 1 & Treasury Secretary, USA \\
	yasuhiro-nakasone & 112 & 1 & 1 & Prime Minister, Japan \\
	paul-volcker & 109 & 1 & 1 & Chairman of Federal Reserve Board, USA \\
	kiichi-miyazawa & 86 & 1 & 1 & Finance Minister, Japan \\
	clayton-yeutter & 85 & 1 & 0 & Trade Representative, USA \\
	nigel-lawson & 66 & 1 & 1 & Chancellor of the Exchequer, UK \\
	dilson-funaro & 58 & 0 & 0 & Finance Minister, Brazil \\
	richard-lyng & 57 & 1 & 0 & Agriculture Secretary, USA \\
	gerhard-stoltenberg & 55 & 1 & 0 & Finance Minister, W. Germany \\
	george-shultz & 50 & 1 & 1 & Secretary of State, USA \\
	margaret-thatcher & 50 & 1 & 1 & Prime Minister, UK \\
	edouard-balladur & 48 & 1 & 1 & Finance Minister, France \\
	james-wright & 47 & 1 & 0 & White House Speaker, Texas Democrat, USA \\
	satoshi-sumita & 44 & 0 & 0 & Bank of Japan Governor, Japan \\
	malcolm-baldrige & 42 & 1 & 0 & Commerce Secretary, USA \\
	marlin-fitzwater & 40 & 1 & 0 & White House Spokesman, USA \\
	alan-greenspan & 39 & 1 & 1 & Chairman of Federal Reserve Board, USA \\
	jaime-ongpin & 36 & 0 & 0 & Finance Secretary, Philippines \\
	jose-sarney & 36 & 1 & 1 & President, Brazil \\
	\botrule
	\end{tabular} \label{tab:top20acount}
}
\end{table}

\begin{table}[ht]
\caption{Top-20 persons in closeness.}
{
	\begin{tabular}{@{}ccccc@{}} 
	\toprule
	person & $cls$ & $H(i)$ & $W(i)$ & notes  \\
	\colrule
	ronald-reagan & 0.136 & 1 & 1 & President, USA \\
	james-baker & 0.120 & 1 & 1 & Treasury Secretary, USA \\
	paul-volcker & 0.114 & 1 & 1 & Chairman of Federal Reserve Board, USA \\
	george-shultz & 0.112 & 1 & 1 & Secretary of State, USA \\
	yasuhiro-nakasone & 0.112 & 1 & 1 & Prime Minister, Japan \\
	kiichi-miyazawa & 0.110 & 1 & 1 & Finance Minister, Japan \\
	gerhard-stoltenberg & 0.109 & 1 & 0 & Finance Minister, W.Germany \\
	nigel-lawson & 0.109 & 1 & 1 & Chancellor of the Exchequer, UK \\
	clayton-yeutter & 0.109 & 1 & 0 & Trade Representative, USA \\
	helmut-kohl & 0.108 & 1 & 1 & Chancellor, W.Germany \\
	edouard-balladur & 0.107 & 1 & 1 & Finance Minister, France \\
	malcolm-baldrige & 0.107 & 1 & 0 & Commerce secretary, USA \\
	marlin-fitzwater & 0.107 & 1 & 0 & White House Spokesman, USA \\
	howard-baker & 0.106 & 1 & 1 & White House Chief of Staff, USA \\
	margaret-thatcher & 0.106 & 1 & 1 & Prime Minister, UK \\
	david-mulford & 0.106 & 1 & 0 & Assistant Treasury Secretary, USA \\
	barber-conable & 0.106 & 1 & 0 & World Bank President \\
	brian-mulroney & 0.106 & 1 & 1 & Prime Minister, Canada \\
	richard-lyng & 0.106 & 1 & 0 & Agriculture Secretary, USA \\
	alan-greenspan & 0.106 & 1 & 1 & Chairman of Federal Reserve Board, USA \\
	\hline
	\end{tabular}
	\label{tab:top20closeness}
}
\end{table}

\subsection{Analysis of Algorithm Performances}

\begin{figure}[htbp]
\centering
\subfigure[] 
{
    \includegraphics[scale=0.7,keepaspectratio=true]{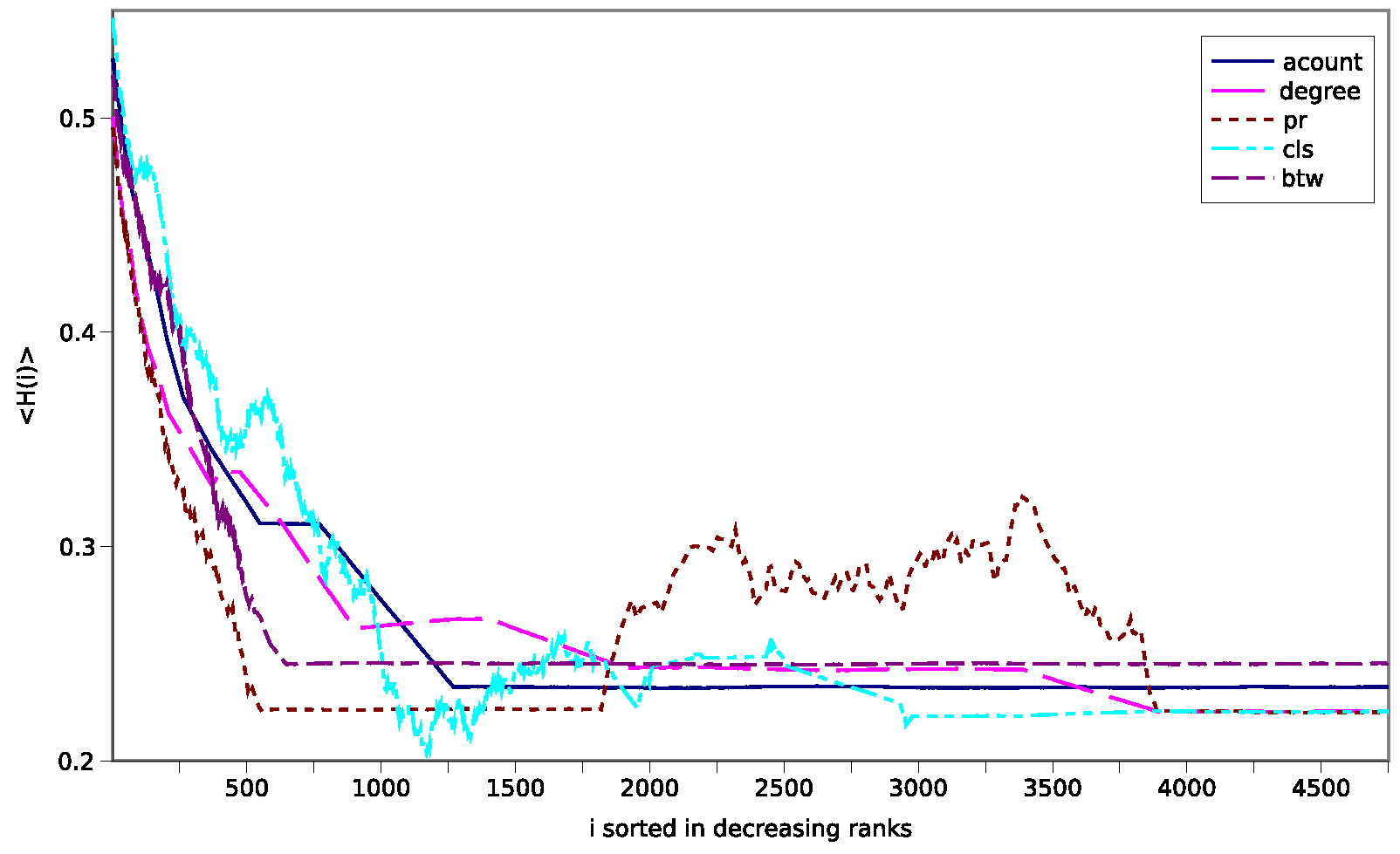}
    \label{fig:mov_avg_hasapage}
}
\vspace{.1in}
\subfigure[] 
{
    \includegraphics[scale=0.7,keepaspectratio=true]{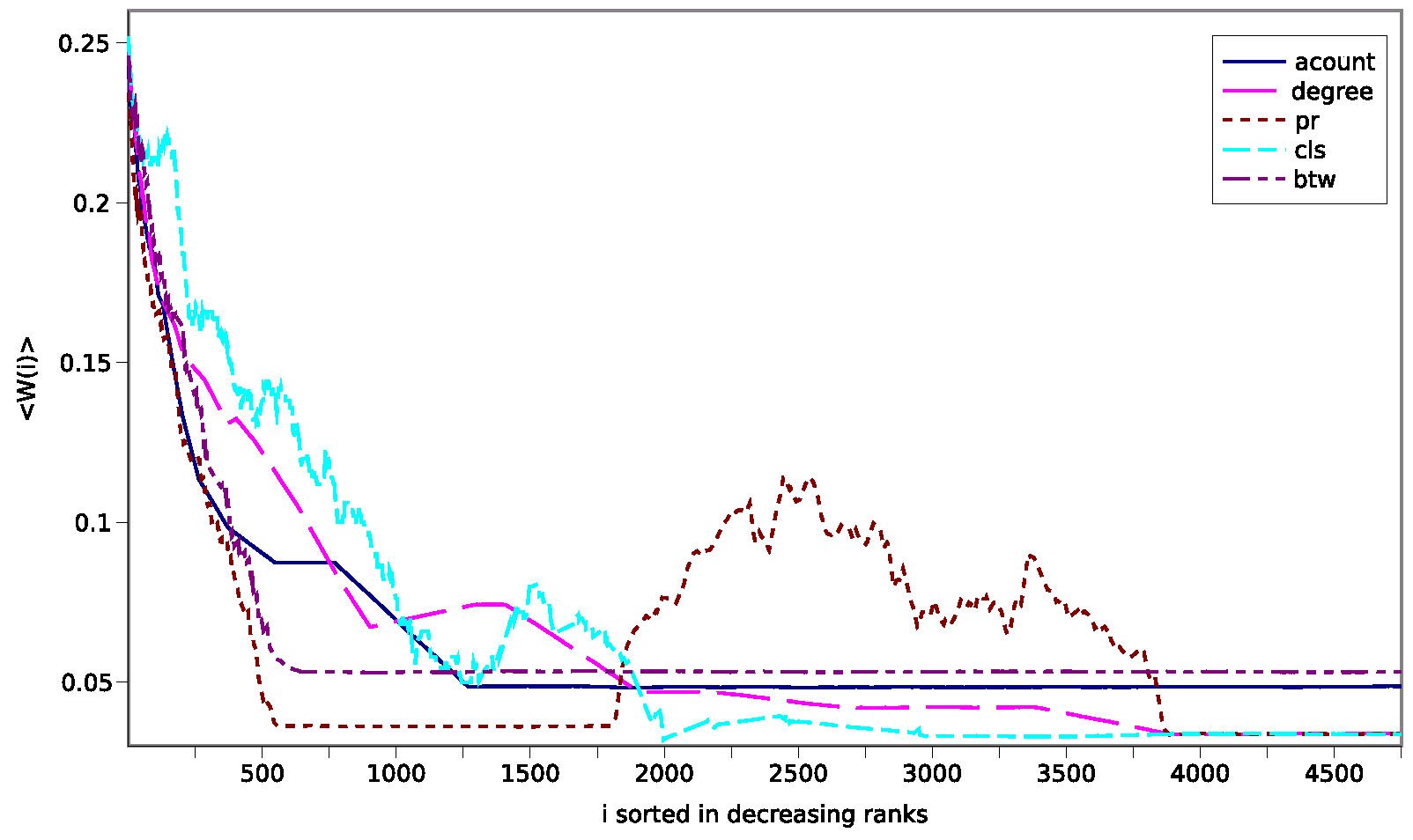}
    \label{fig:mov_avg_iswikified}
}
\caption{(Color online) 
Moving averages of Wikipedia functions with window size 500 for the algorithms
(a) $H(i)$, 
(b) $W(i)$.}
\label{fig:mov_avg} 
\end{figure}

In Fig.~\ref{fig:mov_avg_hasapage} we show the moving averages (using a window size of 500 nodes) of the algorithms for the value of $H(i)$ when nodes are sorted by their ranks in the corresponding algorithms in decreasing order. In producing this graph we randomize the positions of the nodes with equal ranks and take the average value of $H(i)$ corresponding to the position. This produces a near straight line for groups of nodes with equal ranks (e.g. the tails of the plots). The corresponding graph for $W(i)$ is Fig.~\ref{fig:mov_avg_iswikified}.

%

Looking at these graphs we would like an algorithm ideally to put all the positive nodes (i.e. $H(i)=1$ or $W(i)=1$) at the beginning (high ranked positions) and the rest to the tail achieving a separation. 
Given this we see that PageRank and betweenness perform badly as these usually stay below the other algorithms for higher ranked positions. 
At the tail we see betweenness staying above the others which is bad for the same reason that it can not separate positive nodes from the rest.  
Closeness appears to uncover important people earlier and more intensively than others. 

An interesting observation we make on both plots in Fig.~\ref{fig:mov_avg} is the bump PageRank creates close to the middle of the graph with unexpected highs for relatively low PageRanks. 
Investigation reveals that right of this bump are persons who appear in a Reuters article alone with no links thus have very low PageRanks. 
In contrast, to the left of this bump are some overrated persons who also appear in just one article but appear with more than one person resulting in a fully connected sub-graph feeding back each other's ranks. 
This directly relates to the original ``rank sink'' problem in which unreasonably high ranks accumulate in loops without any other outgoing links but have incoming links~\cite{Page1998}. 
This problem was partially remedied by introducing the random jumps 
for the random surfer and removing the nodes with no outgoing links for the rank calculation. 
We can see that the former remedy would not be very effective on this network because the conversion from undirected to directed network creates so many loops in the network. 
Also the latter solution would have no effect in this ranking as those nodes without any links would be place at the tail either way.

\begin{figure}[htbp]
\centering
\subfigure[] 
{
    \includegraphics[scale=0.7,keepaspectratio=true]{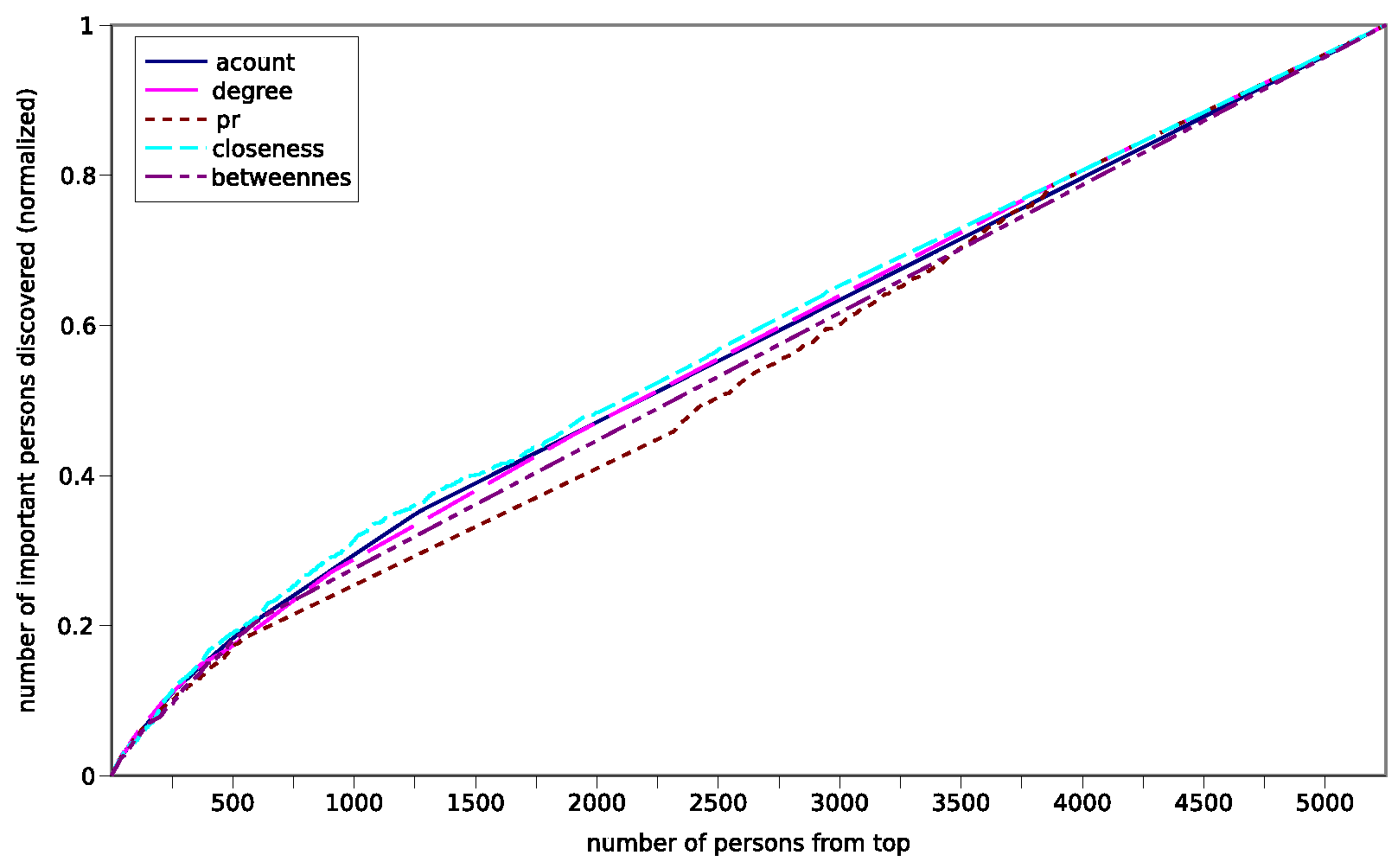}
    \label{fig:cm_diff_hasapage}
}
\vspace{.1in}
\subfigure[] 
{
    \includegraphics[scale=0.7,keepaspectratio=true]{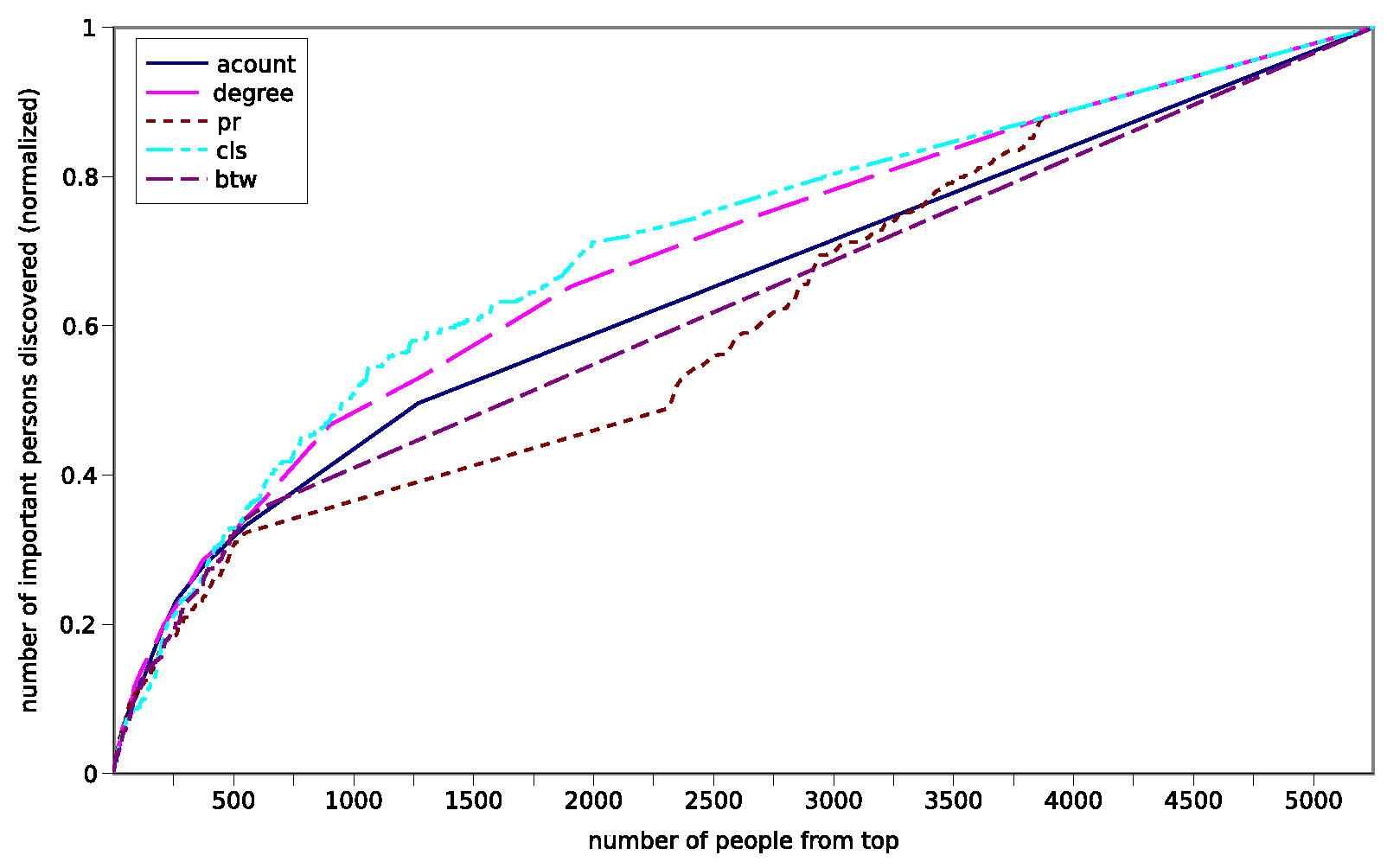}
    \label{fig:cm_diff_iswikified}
}
\caption{(Color online) 
Cumulative performance of algorithms for
(a) $H(i)$, 
(b) $W(i)$.}
\label{fig:cm_diff} 
\end{figure}

The cumulative plots in Fig.~\ref{fig:cm_diff} plot the total number of important people ranked in the top-$t$ people as ranked by an algorithm. 
Similar to the moving average plots we randomize and take averages for nodes with equal ranks. 
We see in these plots that all the algorithms do comparably in the top-500 persons, however later on PageRank and \emph{betweenness} perform badly whereas closeness appears to outperform the others which is consistent with our initial observation above. 
For example in Fig.~\ref{fig:cm_diff_iswikified} we see that if we were to query for the first 1400 persons according to the algorithms, in the results PageRank would have returned about 40\% of all the people with a Wikified page whereas \emph{closeness} would have returned more at about 60\%.

Since the data characteristics of the functions $H(i)$ and $W(i)$ and the rankings produced by the algorithms are different it is not appropriate to do linear correlation. Spearman correlation only requires the orderings produced by two functions to match. Spearman correlation results can be seen on Table~\ref{tab:SpearmanCorrelations}. Note that we have also normalized the results to take into account that maximum (self) correlations for $H(i)$ and $W(i)$ are rather low at 0.597 and 0.203 respectively. We observe that $H(i)$ produces weak (normalized) correlations and $W(i)$ produces medium (normalized) correlations at their highest. The comparative performance we see in these results appear consistent with our observations on the cumulative graphs of Fig.~\ref{fig:cm_diff}.

The low to medium correlations are not unexpected. As we have argued earlier the algorithms are not able to discover all the important persons since that information is not necessarily there, however we should be able to see the differences in their effectiveness in disclosing the evidence that is encoded in the link structures, and this we can observe in the results.

\begin{table}[ht]
\caption{Spearman correlations.}
{
	\begin{tabular}{@{}ccccc@{}} 
	\toprule
	algorithm & $H(i)$ & $H(i)$ (normalized) & $W(i)$ & $W(i)$ (normalized)  \\
	\colrule
	acount & 0.102 & 0.171 & 0.064 & 0.313 \\
	degree & 0.104 & 0.174 & 0.084 & 0.412 \\
	pr & 0.062 & 0.104 & 0.048 & 0.238 \\
	cls & 0.119 & 0.200 & 0.090 & 0.445 \\
	btw & 0.079 & 0.132 & 0.054 & 0.267 \\
	\botrule
	\end{tabular} \label{tab:SpearmanCorrelations}
}
\end{table}

%

\section{Conclusions}
A social network of co-occurrence based on Reuters-21578 corpus is constructed.
Complex Network techniques are used to analyze the network.
It is found to be a small-world network with power-law degree distribution.
The distribution of the component size is governed by a power law.

We reduce the network by taking vertices with minimum degree of 10, where the selection of 10 is arbitrary and 
obtained a network with 49 strongly connected persons. 
This reduction enables us to interpret the results of a community detection algorithm for a large network with unknown community structure. 
The algorithm produced $4$ communities that we can give a meaningful interpretation.

When the network is reduced by removing the edges whose weight is less than 10, where the selection of 10 is also arbitrary, the network become disconnected.  
Interestingly, it becomes almost a star-connected network where Ronald Reagan is in the center 
as a sign of his dominance.
Removal of him would break the already disconnected reduced network and transform to a network where only a few vertices are connected.

The articles have time information. Using this, we have obtained time dependency of co-occurrence. Some people such as Ronald Reagan have activity, consistently. On the other hand, some people have limited time presence due to some particular events.


We try to estimate the importance of the people based on the co-occurrence network of 1987 and compare it with another importance metric of 2007 based on representation in the English Wikipedia, giving 20 years of time difference in between. The time past can make a reliable judgment of who was \emph{not} already important in 1987.
We present a comparative performance analysis for five common centrality measures for our network; article count, degree, PageRank, betweennes and closeness. We found low to medium normalized Spearman correlations between the two metrics. These were in our expected range and in this sense an important contribution of this paper is in providing a means of comparing the relative performances of the algorithms. We found PageRank to be the worst of the five, while closeness was the best performing algorithm. We have discovered an anomaly with PageRank which to some extent can explain why it was performing badly. We believe this may be important as it points out to a pitfall using PageRank in such a network. It is possible to use and combine results from Wikipedias of different languages, as well as different reference sources which we see as a future work.

\section*{Acknowledgments}
This work was partially supported by Bogazici University Research Projects Fund under the grant number 07A105.
Arzucan \"Ozg\"ur thanks Lada Adamic for her helpful comments.



%
%
%
%
%
%
%

\end{document}